\documentstyle[12pt]{article}
\begin{document}
\title{{\bf Is Our Universe Likely to Decay \\
within 20 Billion Years?}
\thanks{Alberta-Thy-08-06, hep-th/0610079}}
\author{
Don N. Page
\thanks{Internet address:
don@phys.ualberta.ca}
\\
Institute for Theoretical Physics\\
Department of Physics, University of Alberta\\
Room 238 CEB, 11322 -- 89 Avenue\\
Edmonton, Alberta, Canada T6G 2G7
}
\date{(2006 October 9)}

\maketitle
\large
\begin{abstract}
\baselineskip 14.5 pt

	Observations that we are highly unlikely to be vacuum
fluctuations suggest that our universe is decaying at a rate faster than
the asymptotic volume growth rate, in order that there not be too many
observers produced by vacuum fluctuations to make our observations
highly atypical.  An asymptotic linear e-folding time of roughly 16 Gyr
(deduced from current measurements of cosmic acceleration) would then
imply that our universe is more likely than not to decay within a time
that is less than 19 Gyr in the future.

\end{abstract}
\normalsize

	Einstein is quoted as saying that the most incomprehensible
thing about the world is that it is comprehensible.  This mystery has
both a philosophical level\footnote{A theistic explanation is that the
universe was created by an omniscient God, so then at least God
comprehends it.  If God made humans in His own image, this might help
explain why humans can also comprehend part of it, though of course we
would like a scientific explanation of the details of how this was
accomplished.} and a scientific level.  The scientific level of the
mystery is the question of how observers within the universe have
ordered observations and thoughts about the universe.

	It seems obvious that our observations and thoughts would be
very unlikely to have the order we experience if we were vacuum
fluctuations, since presumably there are far more quantum states of
disordered observations than of ordered ones.  Therefore, I shall assume
that our observational evidence or order implies that we are not vacuum
fluctuations.

	If we reject solipsism as not the simplest explanation of our
observations, our universe seems to have produced a large number of
varied observers and observations.  Therefore, we cannot expect any
good theory of the universe to predict a unique observer or
observation.  We should instead expect a good theory to predict an
ensemble of observers and observations such that ours is not too
unusual or atypical. (See \cite{SQM,Page-in-Carr} for ways to define
typicality.)  In particular, we should expect a good theory to predict
that ordered observations are not too atypical.  This would not be the
case if the theory predicts that almost all observations arise from
vacuum fluctuations, because only a very tiny fraction of them would be
expected to be ordered (have comprehension).

	If we have a theory for a finite-sized universe that has
ordinary observers of finite size for only a finite period of time
(e.g., during the lifetime of stars and nearby planets where the
ordinary observers evolve), each of which makes only a finite number of
observations (perhaps mostly ordered), then the universe would have
only a finite number of ordinary observers with their largely ordered
observations.  On the other hand, if such a theory predicts that the
universe lasts for an infinite amount of time, then one would expect
from vacuum fluctuations an infinite number of observers (mostly very
short-lived, with very little ordered memory) and observations (mostly
with very little order).  Such a theory would violate the requirement
that a good theory predict our ordered observations as not too unusual
or atypical.  (This argument is a variant of the doomsday argument
\cite{C,Le,N,G,KKP}.)

	Therefore, a good theory for a finite-sized universe should also
predict that it have a finite lifetime.  (For example, this was a
property of the $k=+1$ Friedmann-Robertson-Walker model universes with
nonnegative pressure.)

	For an infinite universe (infinite spatial volume), the
argument is not so clear, since one could get an infinite number of
both ordinary observations and disordered observations, and then there
may be different ways of taking the ratio to say whether either type is
too unusual.  However, here I shall assume that it is appropriate to
take the number of both types of observations per comoving spatial
volume of the universe, which would give the right answer for any
finite universe, no matter how large.  Then we can conclude that any
model universe should not last forever if it has only a finite time
period where ordinary observers dominate \cite{lifetime}.

	The next question is what limits on the lifetime can be deduced
from this argument.  In \cite{lifetime} it was implicitly assumed that
the universe lasted for some definite time $t$ and then ended.  Then
the requirement was that the number of vacuum fluctuation observations
per comoving volume during that time not greatly exceed the number of
ordered observations during the finite time that ordinary observers
exist.  For any power-law expansion with exponent of order unity, I
predicted \cite{lifetime} that the universe would not last past $t \sim
e^{10^{50}}$ years, and for an universe that continues to grow
exponentially with a doubling time of the order of 10 Gyr, I predicted
that the universe would not last past about $10^{60}$ years.

	However, the main point of the present paper is that the
expected lifetime should be much shorter if the universe is expanding
exponentially and just has a certain decay rate for tunneling into
oblivion.  Then the decay rate should be sufficient to prevent the
expectation value of the surviving 4-volume, per comoving 3-volume,
from diverging and leading to an infinite expectation value of vacuum
fluctuation observations per comoving 3-volume.

	Let us take the case in which the decay of the universe proceeds
by the nucleation of a small bubble that then expands at practically the
speed of light, destroying everything within the causal future of the
bubble nucleation event.  Suppose that the bubble nucleation rate, per
4-volume, is $A$ (for annihilation).  Consider the background spacetime
of what the universe would do if it were not destroyed by such an
expanding bubble.  (For simplicity I shall speak as if this background
spacetime had a definite classical 4-geometry, but of course one could
modify the discussion to include quantum uncertainties in its
geometry.)  Then if one takes some event $p$ within this background
spacetime, the probability that the spacetime would have survived to
that event is
\begin{equation}
P(p) = e^{-A V_4(p)},
\label{eq:1}
\end{equation}
where $V_4(p)$ is the spacetime 4-volume to the past of the event $p$ in
the background spacetime.

	Now the requirement that there not be an infinite expectation
value of vacuum fluctuation observations within a finite comoving
3-volume (say region $C$) is the requirement that
\begin{equation}
\int_C P(p) \sqrt{-g} d^4 x < \infty,
\label{eq:2}
\end{equation}
that the total 4-volume within the comoving region, weighted by the
survival probability $P(p)$ for each point, be finite rather than
infinite.

	Let us take the case of an asymptotically de Sitter background
spacetime with cosmological constant $\Lambda$ (and all other matter
decaying away), which asymptotically has a constant logarithmic
expansion rate in all directions of
\begin{equation}
H_\Lambda = \sqrt{\Lambda/3}.
\label{eq:3}
\end{equation}
We assume that there is some sort of Big Bang or beginning of the
universe in the past, so the only question is whether the integral
(\ref{eq:2}) diverges in the asymptotic future part of the background
spacetime. One can then readily calculate that the expectation value of
the 4-volume of the surviving spacetime is finite if and only if
\begin{equation}
A > A_{\mathrm min} = {9\over 4\pi} H_\Lambda^4 = {\Lambda^2\over 4\pi}.
\label{eq:4}
\end{equation}

	If we take $\Omega_\Lambda = 0.72\pm 0.04$ from the third-year
WMAP results of \cite{WMAP} and $H_0 = 72\pm 8$ km/s/Mpc from the Hubble
Space Telescope key project \cite{Freedman}, we get 
\begin{equation}
H_\Lambda = H_0\sqrt{\Omega_\Lambda}
 \approx (16\pm 2\ {\mathrm Gyr})^{-1}
\label{eq:5}
\end{equation}
and therefore
\begin{equation}
A > A_{\mathrm min} \approx (18\pm 2\ {\mathrm Gyr})^{-4}.
\label{eq:6}
\end{equation}

	Let us examine the implications of taking roughly the smallest
possible decay rate consistent with the assumptions and data above,
which is for, say, $\Omega_\Lambda = \Omega_{\Lambda {\mathrm min}} =
0.68$ and $H_0 = H_{0 {\mathrm min}} = 64\ {\mathrm km/s/Mpc} = 0.065\
{\mathrm Gyr}^{-1}$.  This then gives
\begin{equation}
A > A_{\mathrm min}
 = {9\over 4\pi} \Omega_{\Lambda {\mathrm min}}^2 H_{0 {\mathrm min}}^4
 = 6.1\times 10^{-6}\ {\mathrm Gyr}^{-4} = (20\ {\mathrm Gyr})^{-4}
 = 5.2\times 10^{-245} = e^{-562.5},
\label{eq:7}
\end{equation}
where the last two numbers are in Planck units, $\hbar=c=G=1$.

	Now let us apply this to the future of our present universe, to
see how soon it might decay.  Assuming that our universe is spatially
flat and has its energy density dominated by the cosmological constant
and by nonrelativistic matter (e.g., dark matter and baryons), then its
$k=0$ FRW metric may be written as
\begin{equation}
ds^2 = T^2[-d\tau^2 + (\sinh^{4/3}{\tau})(dr^2+r^2d\Omega^2)]
     = a^2(-d\eta^2 + dr^2 + r^2d\Omega^2),
\label{eq:8}
\end{equation}
where $T = 2/(3 H_\Lambda) = 12.4$ Gyr (its maximal value) from the
minimum values for $\Omega_\Lambda$ and $H_0$ given above (which gives
the minimal value of the cosmological constant, $\Lambda = 4/(3T^2)$),
where $\tau = t/T$ is a dimensionless time variable, where the scale
factor is $a = T \sinh^{2/3}{\tau}$, and where
\begin{equation}
\eta = \int_0^\tau {dx \over \sinh^{2/3}{x}}
\label{eq:9}
\end{equation}
is the conformal time.  The present value of $\tau$ is $\tau_0 =
\tanh^{-1}{\sqrt{\Omega_\Lambda}} = 1.17$, again obtaining the numerical
value from using $\Omega_{\Lambda {\mathrm min}}$, and the present value
of the conformal time is $\eta_0 = 3.10$.  (If we had used the mean
$\Omega_{\Lambda} = 0.72$, we would have obtained $\tau_0 = 1.25$ and
$\eta_0 = 3.16$; if we had used the maximum $\Omega_{\Lambda {\mathrm
max}}=0.76$, we would have obtained $\tau_0 = 1.34$ and $\eta_0 =
3.22$.)

	Given this background spacetime, the probability $P$ that it
would have survived to some dimensionless time $\tau$ or corresponding
conformal time $\eta$ is given by Eq. (\ref{eq:1}), where in the metric
(\ref{eq:8}) the spacetime 4-volume to the past of the event $p$ at
conformal time $\eta$ is
\begin{equation}
V_4 = {4\pi\over 3} \int_0^\eta d\eta' a^4(\eta')(\eta-\eta')^3.
\label{eq:10}
\end{equation}
Then in the metric (\ref{eq:8}) this gives the survival probability, as
a function of the dimensionless time $\tau$, as
\begin{equation}
P(\tau) = \exp{\left[-{16\over 27}{A\over A_{\mathrm min}}
           \int_0^\tau dx \sinh^2{x}
            \left(\int_x^\tau {dy\over\sinh^{2/3}{y}}\right)^3\right]}
\label{eq:11}
\end{equation}

	For example, for $\tau = \tau_0 = 1.17$, the minimal value for
today, one gets the maximal survival probability to today (for $A =
A_{\mathrm min}$) as $P(\tau_0) = 0.96$, indicating that there was at
least a 4\% chance that our background universe would have decayed
before the present time.  One can also calculate that the present decay
rate has a minimal value of $-d\ln{P}/dt = 1.04 \times 10^{-11}\
{\mathrm yr}^{-1} = (96\ {\mathrm Gyr})^{-1}$.  With the present earth
population of nearly 7 billion, this would give an minimal expected
death rate of about 7 persons per century.  (Of course, it could not be
7 persons in one century, but all 7 billion with a probability of about
one in a billion per century.)

	If one instead uses the mean measured value $\Omega_\Lambda =
0.72$ and thus $\tau = \tau_0 = 1.25$, one gets the survival probability
to today as $P(\tau_0) < 0.95$, and then with the mean measured value
$H_0 = 72\ {\mathrm km/s/Mpc} = 0.074\ {\mathrm Gyr}^{-1}$ one would get
a present decay rate of $-d\ln{P}/dt > 1.43 \times 10^{-11}\ {\mathrm
yr}^{-1} = (70\ {\mathrm Gyr})^{-1}$.  Going to the approximate maximal
values $\Omega_{\Lambda {\mathrm max}}=0.76$ and $H_0 = H_{0 {\mathrm
max}} = 80\ {\mathrm km/s/Mpc} = 0.082\ {\mathrm Gyr}^{-1}$ would
instead give $P(\tau_0) < 0.95$ and $-d\ln{P}/dt > 1.95 \times 10^{-11}\
{\mathrm yr}^{-1} = (51\ {\mathrm Gyr})^{-1}$.

	However, as the universe is approaching exponential expansion,
the minimal logarithmic decay rate will increase, asymptotically
approaching the logarithmic growth rate of the spatial volume, which has
a minimum value of approximately $3 H_\Lambda = 3 H_{0 {\mathrm min}}
\sqrt{\Omega_{\Lambda {\mathrm min}}} = 16.2\times 10^{-11}\ {\mathrm
yr}^{-1} = (6.2\ {\mathrm Gyr})^{-1}$, about 15.5 times the present
minimum value of the logarithmic decay rate.  Therefore, one cannot
simply use the present decay rate to calculate when the probability of
the survival of the universe will have decreased by a factor of one-half
from the present, what one might call the present half-life of our
universe.

	To calculate an upper limit on the present half-life of our
universe, given that it has lasted until today, we calculate the value
of $\tau = \tau_{1/2}$ for which the minimal rate of decay will lead to
$P(\tau) = P(\tau_0)/2$.  With the minimal values of $\Omega_\Lambda$
and $H_0$ from above, this gives $\tau_{1/2} = 2.71$ and a half-life
(measured from the present until the survival probability is one-half
what it is today) of
\begin{equation}
t_{1/2} < T(\tau_{1/2}-\tau_0) \approx 19.0\ {\mathrm Gyr}. 
\label{eq:12}
\end{equation}
If we instead use the mean values for $\Omega_\Lambda$ and $H_0$ from
above, we get $\tau_{1/2} = 2.72$ and $t_{1/2} \stackrel{<}{\sim}
15.7\ {\mathrm Gyr}$.  The maximal values of $\Omega_\Lambda$ and $H_0$
from above give $\tau_{1/2} = 2.74$ and $t_{1/2} \stackrel{<}{\sim}
13.1\ {\mathrm Gyr}$.

	These calculations give only lower bounds on the annihilation
rate and upper bounds on the half-life of the universe.  It is hard to
give precise upper bounds on the annihilation rate, since even if the
survival probability until today is rather low, we could simply be in
the small fraction of space that does survive.  However, if the
logarithmic decay rate were hundreds of times higher than the minimal
value above and so several inverse Gigayears, it would be highly unusual
for us to have evolved as late as we did in such a rapidly decaying
universe, since there is no known reason why we could not have appeared
on the scene a small number of billion years earlier.  (One could put in
the observed temporal distribution of formation times for
second-generation stars to get a better estimate of how much earlier we
could have appeared in our part of the universe, but I shall leave that
for later publications.)

	As a rather conservative upper limit on the annihilation rate, I
shall here suggest that it cannot be greater than three orders of
magnitude larger than the lower limit of Eq. (\ref{eq:7}), so
\begin{eqnarray}
A_{\mathrm min} = e^{-562.5} < A < A_{\mathrm max}
 \sim 1000A_{\mathrm min}\!
 &=&\! 6.1\times 10^{-3}\ {\mathrm Gyr}^{-4} = (3.6\ {\mathrm Gyr})^{-4}
 \\ \nonumber
 &=&\! 5.2\times 10^{-242} = e^{-555.6}.
\label{eq:13}
\end{eqnarray}

	This upper limit is rather extreme, since if we take the mean
values for $\Omega_\Lambda$ and $H_0$ (not the minimal ones used to
deduce $A_{\mathrm min}$), $A = A_{\mathrm max}$ gives a logarithmic
decay rate today of 8.0 Gyr$^{-1}$ and future half-life of 86 million
years, which would make it surprising that we would live so late.  Thus
this value of $A_{\mathrm max}$ is indeed quite conservatively large. 
$A = A_{\mathrm max}$ also gives a survival probability to today of
less than $4\times 10^{-13}$.  However, this absolute survival
probability is less important than the logarithmic decay rate in making
the time of our appearance within the universe unusual, since we would
presumably need the universe to last a certain number of billions of
years just to be here at all, and that should be put in as one of the
necessary conditions for us to be making our observations of when we
appear.

	We might ask whether it is reasonable that the decay rate of the
universe would be between the two limits above.  In string/M theory, all
that is confidently known for de Sitter spacetime
\cite{Star,GKS,KKLT,Gid,FLW,GM,FS} is that the decay time should be less
than the quantum recurrence time, in Planck units roughly the
exponential of the de Sitter entropy ${\mathcal{S}} = 3\pi/\Lambda$,
\begin{equation}
t_{\mathrm decay} < e^{\mathcal{S}}
 = \exp{\left({\pi\over H_\Lambda^2}\right)}
 = \exp{\left({\pi\over H_0^2 \Omega_\Lambda}\right)}
 \stackrel{<}{\sim} e^{3.7\times 10^{122}} \sim 10^{10^{122.2}}
 \sim e^{e^{282.2}}.
\label{eq:14}
\end{equation}
Of course, this number, which even in Gigayears is enormously greater
than the ten thousand million million millionth power of a googolplex,
is stupendously greater than the decay times I am suggesting.

	In \cite{CDGKL}, Eq. (5.20), a tunneling rate is calculated as
\begin{equation}
A \sim \exp{(-B)} = \exp{\left[-{\pi/4\over m_{3/2}^2}
   {(C-1)^2\over C^2(C-1/2)^2}\right]},
\label{eq:15}
\end{equation}
where $m_{3/2}$ is the gravitino mass, $C^2 > 1$ is a ratio of two
supersymmetric AdS vacua depths, and I have shifted from the normalized
Planck units ($\hbar=c=8\pi G = 1$) used in \cite{CDGKL} to my Planck
units ($\hbar=c=G=1$).

	If we define a renormalized gravitino mass to be
\begin{equation}
\mu_{3/2} \equiv {1-1/(2C)\over (1-1/C)^2} m_{3/2} > m_{3/2},
\label{eq:16}
\end{equation}
which for $C \gg 1$ would be the actual gravitino mass, then $-\ln{A} =
B = \pi/(4\mu_{3/2}^2)$, so the results of Eq. (\ref{eq:13}) imply that
$\mu_{3/2}$ should be within the narrow range (given in Planck units
and then in GeV) of
\begin{equation}
  0.03737 = 4.562 \times 10^{17} {\mathrm GeV} < \mu_{3/2}
< 0.03760 = 4.590 \times 10^{17} {\mathrm GeV},
\label{eq:17}
\end{equation}
of course assuming that this is the correct decay mechanism.

	Looking at this result optimistically, one can say that if this
is the correct decay mechanism, and if the constant $C$ can be
determined, then we would have a fairly precise prediction for the
gravitino mass, with a range of only about 0.6\% width.  Then if the
gravitino were found to have a mass within this range that could be
pinned down even more precisely, we would have a refined estimate for
the decay rate of the universe.  Unfortunately, the predicted mass is
so much larger than what is currently accessible at particle
accelerators that it appears rather discouraging to be able to confirm
or refute this prediction.

	On the pessimistic side, the narrowness of the range for the
gravitino mass in this decay scenario suggests the need for some fine
tuning that appears hard to explain anthropically (i.e., by the
selection effect of observership).  Within a suitable landscape or
other multiverse, this selection effect can explain why the decay rate
is not larger than roughly $A_{\mathrm max}$ (since in those parts of
the multiverse observers would be rare), but I do not see how it can
explain why the decay rate could not be smaller than $A_{\mathrm min}$,
since both a finite number of ordinary observers and an infinite number
of vacuum fluctuation observers per comoving volume could then exist. 
If $A < A_{\mathrm min}$, that would make our observations of very tiny
relative measure, which I would regard as strong evidence against any
theory predicting that result, but it would not be a possibility that
one could rule out just from the requirement that there be observers
and observations.

	Furthermore, if the annihilation rate $A$ can be within the
range above for our part of the multiverse, it would still leave it
unexplained why it is not less than $A_{\mathrm min}$ in some other
part of the multiverse that also allows observers to be produced by
vacuum fluctuations.  If it were less in any such part of the
multiverse, then it would seem that that part would have an infinite
number of vacuum fluctuation observations (almost all of which would be
expected to be much more disordered than ours and so not consistent
with our observations) that is in danger of swamping the ordered
observations in our part (presumably only a finite number per comoving
volume).

	Of course, when one tries comparing the expected number of
observers and observations in different parts of a multiverse, there
are severe problems with comparing the ratios if the total comoving
volumes can be infinite \cite{LM,LLM,GL,Vil,WV,VVW,GV,GSVW,Bousso}.  It
is certainly not so straightforward as comparing various numbers within
the same comoving volume within a single part of the multiverse, as was
done in the analysis of this paper.  Therefore, one cannot be sure that
the potential objection of the previous paragraph is valid, but it is a
worrying note about the results of the present analysis.

	Because of these potential problems with the predictions made
here (that the universe seems likely to decay within 20 billion years),
one might ask how the predictions could be circumvented.

	One obvious idea is that the current acceleration of the
universe is not due to a cosmological constant that would last forever
if the universe itself did not decay away.  Perhaps the current
acceleration is caused by the energy density of a scalar field that is
slowly rolling down a gentle slope of its potential
\cite{Linde,Star,KLP,KL,ASS,KKLLS,GLV,WKLS,Per,lifetime}.  However,
this seems to raise its own issue of fine tuning, since although the
observership selection effect can perhaps explain the small value of
the potential, it does not seem to give any obvious explanation of why
the slope should also be small, unless the scalar field is actually
sitting at the bottom of a potential minimum (which is basically
equivalent to a cosmological constant, modulo the question of whether
one would call it a cosmological constant if it could tunnel away
during the quantum decay of the universe).

	Another possibility is that an infinite number of observers per
comoving volume simply cannot form by vacuum fluctuations, even if the
universe continues to expand forever.  In \cite{lifetime} a possible way
out was given if each observer necessarily spans the entire universe, so
that it cannot be formed by a local quantum fluctuation.  However, this
seems even more far-fetched than the possibility that the dark energy is
slowly decreasing.

	Yet another possibility is that the normalization employed in
this paper to get a finite number of ordinary observers, namely to
restrict to a finite comoving volume, might not be the correct procedure
if our universe really has infinite spatial volume.  However, if our
universe did have finite spatial volume, this procedure would seem
perfectly adequate, so it is not obvious what is wrong with it for the
use I am making of it.  (I do agree that is is problematic for making
comparisons between disconnected parts of the multiverse, since one
would not necessarily know how to compare the sizes of comoving volumes
in the disconnected parts.  However, this problem does not arise for my
comparison of ordinary observers with vacuum fluctuation observers
formed to the future of the same finite comoving volume where the
ordinary observers, namely we, are.)

	Furthermore, there may be a tiny rate (perhaps $\sim
10^{-10^{122.2}}$) for even the comoving future of our spacetime to
tunnel back to an eternal inflationary state and produce an infinite
number of ordinary observers within its future \cite{Vilpriv}.  Then
there would be an infinite number of both ordinary observers and
disordered vacuum fluctuation observers within a finite comoving volume
to the future of our part of spacetime, so again it would become
ambiguous as to which dominates.

	My suggestion then is that one should regularize this infinity
by cutting off the potentially infinite sequence of eternal
inflationary periods and post-inflationary periods (where stars,
planets, and ordinary observers can exist) at some time very far into
one of the post-inflationary periods, but before there is significant
probability that eternal inflation can start again.  Then one would be
led to the predictions of this paper.  However, this is certainly an
{\it ad hoc} proposal, so it might well be wrong.

	So if the predictions made in this paper (that our universe
seems likely to decay within 20 billion years) are wrong, it may be
part of our general lack of understanding of the measure in the
multiverse (or here, even of just different times in the same spacetime
comoving volume).  On the other hand, despite the fine-tuning problems
mentioned above, it is not obvious to me that it really is wrong, so
one might want to take it seriously unless and until some other way is
found to avoid our ordered observations being swamped by disordered
observations from vacuum fluctuations.

	One might ask what the observable effects would be of the decay
of the universe, if ordered observers like us could otherwise survive
for times long in comparison with 20 billion years.

	First of all, the destruction of the universe would occur by a
very thin bubble wall traveling extremely close to the speed of light,
so no one would be able to see it coming to dread the imminent
destruction.  Furthermore, the destruction of all we know (our nearly
flat spacetime, as well as all of its contents of particles and fields)
would happen so fast that there is not likely to be nearly enough time
for any signals of pain to reach your brain.  And no grieving survivors
will be left behind.  So in this way it would be the most humanely
possible execution.

	Furthermore, the whole analysis of quantum cosmology and of
measures on the multiverse seems (at least to me) very difficult to do
without adopting something like the Everett many-worlds version of
quantum theory (perhaps a variant like my own Sensible Quantum
Mechanics or Mindless Sensationalism \cite{SQM,MS}).  Then of course if
there are ``worlds'' (quantum amplitudes) that are destroyed by a
particular bubble, there will always remain other ``worlds'' that
survive.  Therefore, in this picture of the decaying universe, it will
always persist in some fraction of the Everett worlds (better, in some
measure), but it is just that the fraction or measure will decrease
asymptotically toward zero.  This means that there is always some
positive measure for observers to survive until any arbitrarily late
fixed time, so one could never absolutely rule out a decaying universe
by observations at any finite time.

	However, as the measure decreases for our universe to survive
for longer and longer times, a random sampling of observers and
observations by this measure would be increasingly unlikely to pick one
at increasingly late times.  Although observers would still exist then,
they would be increasingly rare and unusual.  Of course, any particular
observer who did find himself or herself there could not rule out the
possibility that he or she is just a very unusual observer, but he or
she would have good statistical grounds for doubting the prediction
made in this paper that he or she really is quite unusual.

	In any case, the decrease in the measure of the universe that I
am predicting here takes such a long time that it should not cause
anyone to worry about it (except perhaps to try to find a solution to
the huge scientific mystery of the measure for the string landscape or
other multiverse theory).  However, it is interesting that the discovery
of the cosmic acceleration \cite{Ries,Perl} may not teach us that the
universe will certainly last much longer than the possible finite
lifetimes of $k=+1$ matter-dominated FRW models previously considered,
but it may instead have the implication that our universe is actually
decaying even faster than what was previously considered.

	I have benefited from email comments by Andrei Linde and Alex
Vilenkin.  This research was supported in part by the Natural Sciences
and Engineering Research Council of Canada.

%\newpage

\end{document}